\documentclass[12pt]{iopart}
\usepackage{iopams}

\begin{document}

\title[Single-particle entanglement]{Single-particle entanglement}

\author{M. Ali Can\dag \footnote[3]{can@fen.bilkent.edu.tr}, Alexander Klyachko\ddag and
Alexander Shumovsky\dag  }

\address{\dag\ Department of Physics, Bilkent University, Bilkent,
Ankara 06533, Turkey}
\address{\ddag\ Department of Mathematics, Bilkent University,
Bilkent, Ankara 06533, Turkey}


\begin{abstract}
Using the approach to quantum entanglement based on the quantum
fluctuations of observables, we show the existence of perfect
entangled states of a single ``spin-$1$" particle. We give
physical examples related to the photons, condensed matter
physics, and particle physics.
\end{abstract}



\submitto{accelerated publication in Journal of Optics B}

\maketitle

\newpage


In the usual treatment, the quantum entanglement is associated
with the specific {\it nonlocal} correlations among the parts of a
quantum system that has no classical analog (e.g., see [1]). This
assumes that the entangled system should consist of two or more
parts. At the same time, there is a strong interest in the
single-particle (especially single-photon) entanglement [2-8]. In
particular, the possibility to use single-photon entanglement in
quantum cryptography has been discussed recently [7].

The single-photon entanglement is usually considered in terms of
the two-qubit entanglement. One of qubits is intrinsic property of
the photon like polarization, while the second qubit corresponds
to the spatial degrees of freedom, defined by the two spatial
modes of a single photon. These modes can be produced either by a
beam splitter [2,3,8] or through the use of two identical
cavities, containing single excitation [6].

Undoubtedly, it seems to be of high interest to consider
entanglement caused only by the intrinsic degrees of freedom of a
single particle.

Here we examine the single-particle entanglement
from  the perspective  of recent approach, treating entanglement
as a
manifestation of quantum fluctuations in a state where the
fluctuations come to their extreme
[9-13]. In particular, it was shown that the {\it completely
entangled} (CE) states of a given system can be defined in terms
of a certain variational principle for the total amount of quantum
fluctuations [12]. It should be stressed that every entangled
state can be transformed into CE state by a $SLOCC$ transformation
[14-16] that can change the amount of entanglement but can't
neither create nor destroy it. Mathematically $SLOCC$
transformation amounts to action of {\it complexified\,} dynamic
group $G^c$.

The essence of the approach can be formulated as follows
[11,12,17]. Let $\mathbb{H}_A$ be the space of states of a quantum
system $A$, and $\mathcal{L}$ be the Lie algebra generated by
observables we are going to measure in the course of experiment,
or, what is the same, by the Hamiltonians available for
manipulation with quantum states. $\mathcal{L}$ is said to be the
Lie algebra of {\it essential observables\,}, and the
corresponding compact group $G=\exp(\mathcal{L})$ is called the
{\it dynamic group\,} of the system $A$. For example, for a two
component system $\mathbb{H}_{AB}=\mathbb{H}_A\otimes
\mathbb{H}_B$ with full access to local degrees of freedom the
dynamic group is
$\mbox{SU}(\mathbb{H}_A)\times\mbox{SU}(\mathbb{H}_B)$. The
corresponding group of $SLOCC$ transformations
$G^c=\exp(\mathcal{L}^c)$ is defined by complexified algebra
$\mathcal{L}^c=\mathcal{L}\otimes \mathbb{C}$. In the above
example,
$G^c=\mbox{SL}(\mathbb{H}_A)\times\mbox{SL}(\mathbb{H}_B)$.

The key physical quantity responsible for entanglement  of a state
$\psi \in\mathbb{H}_A$ is its {\it total variation\,}
\begin{eqnarray}
V_{tot}(\psi)= \sum_i (\langle \psi|\mathcal{O}_i^2|\psi \rangle -
\langle \psi|\mathcal{O}_i|\psi \rangle^2), \label{fluct}
\end{eqnarray}
where summation is performed over an orthonormal basis
$\mathcal{O}_i$ of the Lie algebra of essential observables
$\mathcal{L}$. The crucial point is that this quantity is
independent of the basis $\mathcal{O}_i$, and reflects the total
{\it amount of quantum fluctuations\,} of the system in the state
$\psi$. For spin group $\mbox{SU}(2)$, one can use spin projection
operators $S_x,S_y,S_z$ as the basis of $\mathcal{L}=su(2)$.

The quantity (1) bears a similarity with the so-called {\it skew
information} that has been introduced by Wigner [18,19] to specify
the amount of information, carrying by a quantum state with
respect to noncommuting observables, whose measurement needs
macroscopic apparatuses. In turn, the observables associated with
the additive conserved quantities like energy can be measured with
microscopic apparatuses. The main difference between our approach
and that by Wigner consists in the definition of fundamental
observables in terms of the dynamic symmetry of the system.

To clarify the physical meaning of (1), note that in the case of
classical observables represented by c-numbers the total amount of
fluctuations is equal to zero. Thus, the nonzero value of (1)
specifies the {\it remoteness} of the state $\psi$ from the
``classical reality", i.e., from the result of classical
measurements.

CE states $\psi_{CE} \in \mathbb{H}_S$
have the following {\it extremality property\,} [12]
\begin{eqnarray}
V_{tot}(\psi_{CE})= \max_{\psi \in \mathbb{H}_S} V_{tot}(\psi).
\label{2}
\end{eqnarray}
This means that CE states provide the maximal amount of quantum
fluctuations in the system. In other words, CE states are
maximally remote from the ``classical reality". This clarifies the
fact that entanglement has no classical analog.

On the contrary, generalized coherent states correspond to the
minimal amount of quantum fluctuations [17] (concerning
generalized coherent states, see Ref. [20]). Thus, they are
closest to the ``classical reality".

Equation (\ref{2}) plays in  entanglement the same role  as
variational principles in mechanics. Using differential criterion
of extremum, one can recast it into the form
\begin{equation}\label{aver0}
\langle \psi_{CE}\mid \mathcal{O}\mid \psi_{CE}\rangle=0,\qquad
\forall \mathcal{O}\in \mathcal{L},
\end{equation}
which tells that in CE state the system is in the center of its
quantum fluctuations. The definition (\ref{aver0}) does not assume
the nonlocality of the system A, and therefore can be used to
study entanglement in the single-component systems.

As an example of some practical interest consider spin-1 system
with dynamic group $\mbox{SU}(2)$ in its three-dimensional
irreducible representation $\mathbb{H}_1$. It can be realized as a
dipole photon with total angular momentum $j=1$ [21,22]. Another
realization is provided by the superfluid $^3\mbox{He}$, where
both spin and orbital momenta of a Cooper pair are equal to one
[23,24]. In the most common B-phase of $^3\mbox{He}$, spin and
orbital momenta of a Cooper pair are completely entangled, while
in other stable phases they are separated, and can be treated as
independent spin-1 states. Say, in A-phase, spin part of Cooper
pair is completely entangled, while its orbital part is coherent.
For $\beta$-phase, the situation is just the opposite: spin part
is coherent and orbital is entangled. In the so called polar
phase, both spin and orbital parts are entangled spin-1 states. In
exotic $A_1$-phase, both components are coherent. These are all
stable phases, representing local minima of free energy.

To clarify the structure of CE states in a single spin-1 system we
start with Clebsch-Gordon decomposition
\begin{eqnarray}
\mathbb{H}_{\frac{1}{2}} \otimes
\mathbb{H}_{\frac{1}{2}} = \mathbb{H}_1 \oplus \mathbb{H}_0,
\label{3}
\end{eqnarray}
of two spin-$\frac{1}{2}$ systems into symmetric component
$\mathbb{H}_1$ of spin 1, and skew symmetric scalar component
$\mathbb{H}_0$. If we denote the base states in
$\mathbb{H}_{\frac{1}{2}}$ by $|\uparrow \rangle$ and $|\downarrow
\rangle$, then the basis of $\mathbb{H}_1$ is represented by the
symmetric triplet
\begin{eqnarray}
\left\{ \begin{array}{l} |\uparrow\uparrow \rangle \\
|\downarrow\downarrow \rangle \\ \frac{1}{\sqrt{2}}
(|\uparrow\downarrow \rangle +|\downarrow\uparrow \rangle )
\end{array} \right. \label{triplet}
\end{eqnarray}
while the antisymmetric singlet
\begin{eqnarray}
\frac{1}{\sqrt{2}} (|\uparrow\downarrow \rangle
-|\downarrow\uparrow \rangle ) \label{singlet}
\end{eqnarray}
corresponds to $\mathbb{H}_0$. Since the states of spin-1 system
under consideration can always be specified by the projection of
spin onto the quantization axis $|m \rangle$, the states
(\ref{triplet}) can be interpreted as the states $|m=1 \rangle$,
$|m=-1 \rangle$, and $|m=0 \rangle$, respectively. From the
physical point of view, this means that if a single spin-1 system,
prepared initially in the state $|m=0 \rangle$, decays into the
two spin-$\frac{1}{2}$ objects, they should be observed in the EPR
(Einstein-Podolsky-Rosen) state (the last state in
(\ref{triplet})). This is an indication that spin-1 state $|m=0
\rangle$ is entangled. The other two states $|m=\pm1 \rangle$ in
the triplet (\ref{triplet}) are coherent and decay into
disentangled spin-$\frac12$ components.

To classify spin-1 states, it is convenient to treat
$\mathbb{H}_1$ as complexification of three-dimensional Euclidean
space
\begin{equation}\label{euclid}
\mathbb{H}_1=\mathbb{E}^3 \otimes \mathbb{C}
\end{equation}
with dynamical symmetry group $\mbox{SU}(2)\approx \mbox{SO}(3)$,
acting by rotations in $\mathbb{E}^3$. Then, every state $|\psi
\rangle$ in $\mathbb{H}_1$ can be represented as the
complex superposition
\begin{eqnarray}
|\psi \rangle = \cos \varphi \cdot |\vec{\mu} \rangle +i \sin
\varphi
\cdot |\vec{\nu} \rangle ,\qquad 0 \leq \varphi
\leq
\pi/4, \label{psi}
\end{eqnarray}
of two orthonormal vectors $\vec{\mu},\vec\nu\in\mathbb{E}^3$.
Note that one orthonormal pair  $\vec{\mu},\vec{\nu}\in
\mathbb{E}^3$ can be transformed into another by a rotation.
Hence, the angle $\varphi$ is the unique {\it intrinsic\,}
invariant of spin-1 state. Therefore, it is not surprising that
its measure of entanglement can be expressed via $\varphi$.

We'll see later that $\varphi =0$ corresponds to the CE states,
while $\varphi = \pi/4$ gives unentangled (coherent) states. In
theory of superfluid $^3\mbox{He}$, the former are known as the
{\it unitary\,} states.

Spin projection operator $S_{\vec{\omega}}$ onto direction
$\vec{\omega}\in \mathbb{E}^3$ in representation (\ref{euclid})
amounts to infinitesimal rotation with angular velocity
$\vec{\omega}$ given by the cross product
\begin{equation}\label{spin_proj}
S_{\vec{\omega}}: x\mapsto i\vec{\omega} \times \vec{x},\quad
\vec{x}\in \mathbb{E}^3.
\end{equation}
Hence, $S_{\vec{\nu}}|\vec{\nu}\rangle=0$, i.e.
$|\vec{\nu}\rangle$ is a state with zero spin projection onto
direction $\vec{\nu}$. Moreover, by (\ref{spin_proj})
\begin{equation}
\langle\vec{\nu}\mid
S_{\vec{\omega}}\mid\vec{\nu}\rangle=(\vec{\nu},\vec{\omega},\vec{\nu})=0,\quad
\forall \vec{\omega}\in \mathbb{E}^3
\end{equation}
and by criterion (\ref{aver0}), $|\vec{\nu}\rangle$ is CE state.
For the general state (\ref{psi}), we get
\begin{equation}
\langle\psi\mid
S_{\vec{\omega}}\mid\psi\rangle=2\sin\varphi\cos\varphi
(\vec{\mu},\vec{\omega},\vec{\nu})=\sin(2\varphi)(\vec{\mu},\vec{\omega},\vec{\nu}).
\end{equation}
Hence, $|\psi\rangle$ is the CE state only for $\varphi=0$. So, we
arrive at characterization of CE states as those with spin
projection $m=0$ onto some direction. Typical examples are the
states
\begin{eqnarray}
\left\{ \begin{array}{lcl} |\psi_0 \rangle & = & |0 \rangle \\
|\psi_{\pm 1} \rangle & = & \frac{1}{\sqrt{2}} (|+1 \rangle \pm
|-1 \rangle ) \end{array} \right. \label{basis}
\end{eqnarray}
which form a completely entangled basis in $\mathbb{H}_1$. One of
those states $|\psi_{0} \rangle =|0 \rangle$ formally corresponds
to the EPR state in (\ref{triplet}).

Taking into account that the general state (\ref{psi}) of the
spin-1 system can be formally represented in the form of the
two-qubit state
\begin{eqnarray}
|\psi \rangle = \psi_{\uparrow\uparrow} |\uparrow\uparrow \rangle
+ \psi_{\downarrow\downarrow}|\downarrow\downarrow \rangle +
\psi_{\uparrow\downarrow}(|\uparrow\downarrow \rangle
+|\downarrow\uparrow \rangle ) \nonumber
\end{eqnarray}
in the symmetric sector, and that the {\it concurrence} (measure
of entanglement in the case of two qubits [25]) has the form
\begin{eqnarray}
\mathcal{C}(\psi)=2|\det
[\psi]|=2|\psi_{\uparrow\uparrow}\psi_{\downarrow\downarrow}-
\psi_{\uparrow\downarrow}\psi_{\downarrow\uparrow}|, \nonumber
\end{eqnarray}
we can conclude that the amount of entanglement in CE basis
(\ref{basis}) can be measured by the expression
\begin{eqnarray}
\mathcal{C}(\psi)=2|\psi_{+1} \psi_{-1}- \psi_0^2/2|,
\label{concur}
\end{eqnarray}
which represents the concurrence in the case of spin-1 system. It
is interesting that the concurrence can also be expressed in terms
of the total amount of fluctuations (\ref{fluct}) as follows
\begin{eqnarray}
\mathcal{C}(\psi)=
\sqrt{\frac{V_{tot}(\psi)-V_{min}}{V_{max}-V_{min}}} . \nonumber
\end{eqnarray}
In terms of the inherent parameter $\varphi$ introduced by Eq.
(\ref{psi}), the concurrence (\ref{concur}) takes the form
\begin{eqnarray}
\mathcal{C}(\psi)= \cos 2 \varphi, \quad \varphi \in [0, \pi /4].
\nonumber
\end{eqnarray}
Similar analysis can also be done in the case of mixed states of a
single spin-1 system.

Concerning physical realizations, let us mention first that the
three-dimensional entanglement in orbital angular momentum of
photons [26,27] provides an example, illustrating the above
theory. Namely, a single photon in Laguerre-Gauss beam in the
state $|m=0 \rangle$ is entangled by itself. Let us stress that in
the usual treatment, entanglement with respect to the orbital
angular momentum of a pair of photons [26,27] is discussed.

According to our result, a single dipole photon [22] with angular
momentum $j=1$ and projection $m=0$ is always in the CE state. In
view of the above interpretation, we can assume that such a photon
may decay into a pair of entangled particles. In other words, the
electron-positron pair created by the photodecay of the dipole
photon with $m=0$ should be prepared in the CE EPR state (the last
state in (\ref{triplet})) with respect to the spin of charged
particles. This may be observed in the presence of a strong
electric field, which separates the particles with opposite charge
and, unlike the magnetic field, does not influence the spin state.
Other photon decay processes such as resonance down-conversion and
Raman scattering with creation of the entangled pairs can also be
described using the above formalism.

The examples of spin-1 entangled states of Cooper pairs in
superfluid $^3He$ were mentioned above.

Another example of single-particle CE state is provided by the
isodoublet of quarks with only two flavors, namely up- and
down-quarks, forming $\pi$-mesons [28]. The $\pi^{\pm}$-mesons
represent the coherent states with respect to the quarks
\begin{eqnarray}
\pi^+=u\bar{d}, \quad \pi^-= \bar{u}d . \nonumber
\end{eqnarray}
In contrast, $\pi^0$-meson is prepared in the CE state of the type
of $|\psi_0 \rangle$ in (\ref{basis})
\begin{eqnarray}
\pi^0 = \frac{u\bar{u}-d\bar{d}}{\sqrt{2}} . \nonumber
\end{eqnarray}
Since CE corresponds to the maximum of the total amount of
fluctuations, all one can expect is that $\pi^0$ meson should be
less stable than $\pi^{\pm}$. In fact, the experimental ratio of
the lifetimes is $\tau_0 / \tau_{\pm} \sim 10^{-9}$ [28].

Thus, we have shown that the single-particle spin-1 system
prepared in the state with spin projection $m=0$ always manifest
complete entanglement, defined in terms of the maximum total
amount of quantum fluctuations. This means that those states are
less stable than not CE states, and that the possible decay of
those states leads to creation of EPR pairs. The above
consideration shows that the notion of the single-particle
entanglement as well as the approach used for its description are
quite general. In particular, they can be used for analysis of
states of photons, quantum liquids, and elementary particles.

\end{document}